\documentclass[letter]{sig-alternate-2013}
\usepackage{xspace}
\usepackage{times}
\usepackage{float}
\usepackage{multirow}
\usepackage{color}
\usepackage{cite}
\usepackage{url}
\usepackage{textcomp}
\usepackage{alltt}
\usepackage{upquote}
\usepackage{listings}
\usepackage{xcolor}
\usepackage{pdflscape}
\usepackage{multicol}
\usepackage{epsfig}
\usepackage{epstopdf}
\usepackage{graphicx}
\usepackage{subfigure}
\usepackage{relsize}
\usepackage{comment}
\usepackage{caption}
\usepackage{alltt}
\usepackage{verbdef}

\usepackage{enumitem}
\usepackage[hidelinks]{hyperref}

\setlist{itemsep=-0.1em,topsep=-0.03em}

\lstdefinelanguage{JavaScript}{
  keywords={typeof, new, true, false, try, catch, function, return, null, switch, var, if, in, while, do, else, case, break},
  keywordstyle=\color{blue}\bfseries,
  ndkeywords={class, export, boolean, throw, implements, import, this},
  ndkeywordstyle=\color{darkgray}\bfseries,
  identifierstyle=\color{black},
  sensitive=false,
  comment=[l]{//},
  morecomment=[s]{/*}{*/},
  commentstyle=\color{purple}\ttfamily,
  stringstyle=\color{red}\ttfamily,
  morestring=[b]"
}

\newfont{\mycrnotice}{ptmr8t at 7pt}
\newfont{\myconfname}{ptmri8t at 7pt}
\let\crnotice\mycrnotice%
\let\confname\myconfname%

\permission{Permission to make digital or hard copies of all or part of this work for personal or classroom use is granted without fee provided that copies are not made or distributed for profit or commercial advantage and that copies bear this notice and the full citation on the first page. Copyrights for components of this work owned by others than the author(s) must be honored. Abstracting with credit is permitted. To copy otherwise, or republish, to post on servers or to redistribute to lists, requires prior specific permission and/or a fee. Request permissions from permissions@acm.org.}
\conferenceinfo{ASIA CCS'14,}{June 4--6, 2014, Kyoto, Japan. \\
{\mycrnotice{Copyright is held by the owner/author(s). Publication rights licensed to ACM.}}}
\copyrightetc{ACM \the\acmcopyr} 
\crdata{978-1-4503-2800-5/14/06\ ...\$15.00.\\
http://dx.doi.org/10.1145/2590296.2590324
}

\toappear{Pre-print. \\To appear in the ACM ASIA CCS'14, June 4--6, 2014, Kyoto, Japan\\http://dx.doi.org/10.1145/2590296.2590324}

\begin{document}
%
% --- Author Metadata here ---
%\conferenceinfo{ASIACCS '14,}{June 4--6, 2014, Kyoto, Japan.}
%\CopyrightYear{2014} % Allows default copyright year (20XX) to be over-ridden - IF NEED BE.
%\crdata{0-89791-88-6/97/05}  % Allows default copyright data (0-89791-88-6/97/05) to be over-ridden - IF NEED BE.
% --- End of Author Metadata ---

\title{Scanning of Real-world Web Applications for Parameter Tampering Vulnerabilities}
%\subtitle{[Extended Abstract]
%\titlenote{A full version of this paper is available as
%\textit{Author's Guide to Preparing ACM SIG Proceedings Using
%\LaTeX$2_\epsilon$\ and BibTeX} at
%\texttt{www.acm.org/eaddress.htm}}}
%
% You need the command \numberofauthors to handle the 'placement
% and alignment' of the authors beneath the title.
%
% For aesthetic reasons, we recommend 'three authors at a time'
% i.e. three 'name/affiliation blocks' be placed beneath the title.
%
% NOTE: You are NOT restricted in how many 'rows' of
% "name/affiliations" may appear. We just ask that you restrict
% the number of 'columns' to three.
%
% Because of the available 'opening page real-estate'
% we ask you to refrain from putting more than six authors
% (two rows with three columns) beneath the article title.
% More than six makes the first-page appear very cluttered indeed.
%
% Use the \alignauthor commands to handle the names
% and affiliations for an 'aesthetic maximum' of six authors.
% Add names, affiliations, addresses for
% the seventh etc. author(s) as the argument for the
% \additionalauthors command.
% These 'additional authors' will be output/set for you
% without further effort on your part as the last section in
% the body of your article BEFORE References or any Appendices.

\numberofauthors{1} %  in this sample file, there are a *total*
% of EIGHT authors. SIX appear on the 'first-page' (for formatting
% reasons) and the remaining two appear in the \additionalauthors section.
%

\author{
\alignauthor Adonis P.H. Fung{\large \footnotemark[2] }, Tielei Wang{\large \footnotemark[3] }, K.W. Cheung{\large \footnotemark[2] }, and T.Y. Wong{\large \footnotemark[2]}\\
\affaddr{{\footnotemark[2] } The Chinese University of Hong Kong, Shatin, NT, Hong Kong} \\
\affaddr{{\footnotemark[3] } Georgia Institute of Technology, Atlanta, GA, USA} \\
\email{phfung@ie.cuhk.edu.hk,tielei@gatech.edu,\{kwcheung@ie,tywong@cse\}.cuhk.edu.hk}
}

% \author{
% You can go ahead and credit any number of authors here,
% e.g. one 'row of three' or two rows (consisting of one row of three
% and a second row of one, two or three).
%
% The command \alignauthor (no curly braces needed) should
% precede each author name, affiliation/snail-mail address and
% e-mail address. Additionally, tag each line of
% affiliation/address with \affaddr, and tag the
% e-mail address with \email.
%

% \alignauthor
% Adonis P.H. Fung\\
      % \affaddr{The Chinese University of Hong Kong}\\
      % \email{phfung@ie.cuhk.edu.hk}

% \alignauthor
% Tielei Wang\\
       % \affaddr{Georgia Institute of Technology}\\
       % \email{tielei@gatech.edu}

% \alignauthor K.W. Cheung and T.Y. Wong
       % \affaddr{The Chinese University of Hong Kong}\\
       % \email{{kwcheung@ie,tywong@cse}.cuhk.edu.hk}
% }
% There's nothing stopping you putting the seventh, eighth, etc.
% author on the opening page (as the 'third row') but we ask,
% for aesthetic reasons that you place these 'additional authors'
% in the \additional authors block, viz.
% \additionalauthors{Additional authors: John Smith (The Th{\o}rv{\"a}ld Group,
% email: {\texttt{jsmith@affiliation.org}}) and Julius P.~Kumquat
% (The Kumquat Consortium, email: {\texttt{jpkumquat@consortium.net}}).}
% \date{30 July 1999}
% Just remember to make sure that the TOTAL number of authors
% is the number that will appear on the first page PLUS the
% number that will appear in the \additionalauthors section.

\maketitle

\begin{abstract}

%For instance, online banking transfer applications typically adopt a common workflow that involves a page to initiate a transfer request, another to review it, and finally a page for acknowledgment. The review page is thus literally a repeater that forwards user-supplied parameters. 

Web applications require exchanging parameters between a client and a server to function properly. In real-world systems such as online banking transfer, traversing multiple pages with parameters contributed by both the user and server is a must, and hence the applications have to enforce workflow and parameter dependency controls across multiple requests. An application that applies insufficient server-side input validations is however vulnerable to parameter tampering attacks, which manipulate the exchanged parameters. Existing fuzzing-based scanning approaches however neglected these important controls, and this caused their fuzzing requests to be dropped before they can reach any vulnerable code. 

In this paper, we propose a novel approach to identify the workflow and parameter dependent constraints, which are then maintained and leveraged for automatic detection of server acceptances during fuzzing. We realized the approach by building a generic blackbox parameter tampering scanner. It successfully uncovered a number of severe vulnerabilities, including one from the largest multi-national banking website, which other scanners miss.

%honors the parameter dependency while fuzzing, which is essential to uncovering parameter tampering vulnerabilities. We realize this by building a generic blackbox parameter tampering scanner that can preserve the intended workflow as enforced by real-world applications. 
%The importance of this work is demonstrated by the vulnerabilities uncovered in real-world applications including banks.

\end{abstract}

%\category{K.6.5}{Security and Protection}{Unauthorized access}
%\terms{Theory}

\section{Introduction}
\label{sec:crs_intro}

\begin{figure}[ht*]
\centering
\includegraphics[width=0.485\textwidth]{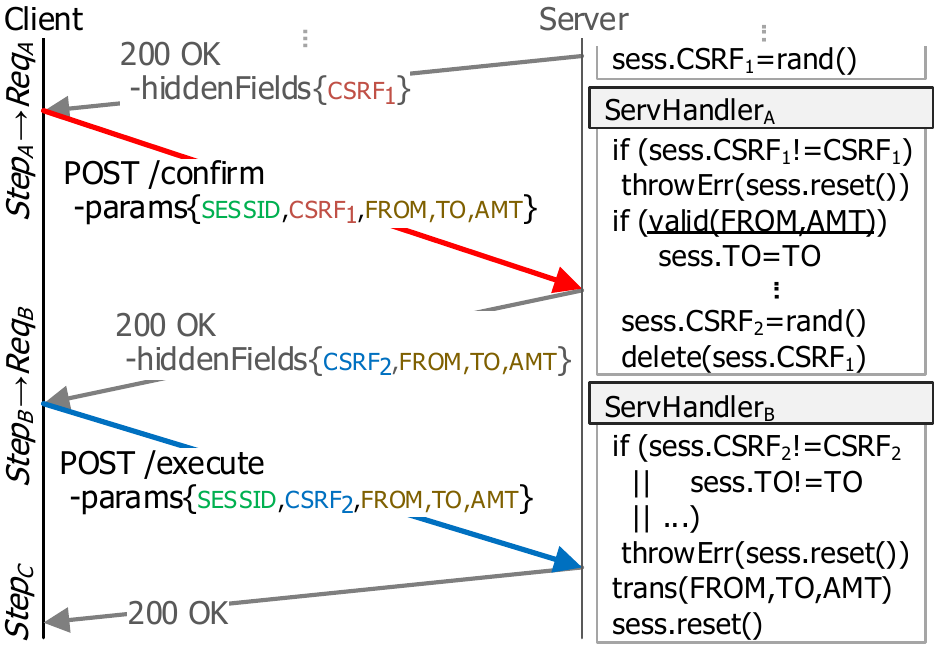}
\caption{A vulnerable banking transfer application enforces workflow and parameter dependency controls across multiple requests}
\label{fig:runningexample}
\vspace{-1.5em}
\end{figure}

Web applications typically require traversing multiple pages with parameters exchanged between a client and a server to complete even a single action. 
Figure~\ref{fig:runningexample} depicts a simplified workflow of an online banking transfer application used by the HSBC bank, in which a user can transfer a certain amount of money (i.e., \verb"AMT") from his account \verb"FROM" only to an authorized account \verb"TO". While user-supplied transaction details (i.e.,\verb"FROM", \verb"TO", \verb"AMT") are taken from input elements of a form, those server-generated ones such as session ID (\verb"SESSID") and the one-time use tokens working against Cross-Site Request Forgeries (CSRF$_{1\&2}$) are respectively set in Cookies and hidden fields. In Step$_A$, client-side validations are applied to restrict and instantly prompt the user for input corrections before the form can be submitted. Upon a valid submission, the ServHandler$_A$ verifies the CSRF$_1$ token, validates the user inputs, and responds with a review page for Step$_B$. The user confirms the transaction by submitting Req$_B$, and finally the ServHandler$_B$ executes the actual banking transfer and returns an acknowledgment page for user's reference at Step$_C$. It is worth noting that ServHandler$_B$ enforces, among others, that the received token CSRF$_2$ and \verb"TO" account respectively match with their previously stored values (i.e., \verb"sess.CSRF"$_2$ and \verb"sess.TO").

Such kind of web applications are however vulnerable to parameter tampering that is known to attack insecure direct object references, which is ranked No. 4 in the OWASP Top 10 Web Application Security Risks~\cite{owasp-insecure-direct-object}. The vulnerability often arises from a misconception that the object references directly exposed as parameters to the client-side (and their validations) are assumed immutable, and thus improper or insufficient validations were applied at the server~\cite{immutable}. As shown in the motivating example, the server does not validate whether the \verb"TO" account is authorized (i.e., \verb"TO" is missing at \verb"valid(FROM,AMT)" in ServHandler$_A$). Hence, an attacker who has compromised a victim's session can tamper the \verb"TO" parameter and bypass any associated client-side validations in order to commit unauthorized banking transfers.

The parameter tampering vulnerability revealed above is however hardly discoverable despite many research efforts that were dedicated to web vulnerability scanning~\cite{waptec,notamper,state-aware,swift,ripley,kudzu,logicvul,parampollution}. There are two fundamental reasons. First, blackbox fuzzing-based scanners such as~\cite{notamper,acunetix,webscarab,tamperdata} cannot preserve the intended workflow. It is because they literally work in a ``crawl-once-fuzz-many'' manner, that captures a set of requests only once during crawling, and these requests will become the only bases to generate subsequent fuzzing requests. However, the one-time use tokens such as CSRF$_{1\&2}$ expire as soon as they are accepted by the server during crawling. Regardless of how parameters are mutated, subsequent fuzzing attempts that reuse such expired tokens will all be rejected.

Second, existing blackbox scanners neglect cross-request parameter dependencies. To uncover the vulnerbility in the banking example, the \verb"TO" account from Req$_A$ and the one later sent in Req$_B$ must be equal. It is because ServHandler$_A$ stores the user-supplied parameter \verb"TO" from Req$_A$ into \verb"sess.TO". When receiving Req$_B$, ServHandler$_B$ enforces that \verb"TO" in Req$_B$ must match with the stored \verb"sess.TO". If they are different, Req$_B$ will be rejected before any parameters can reach the vulnerable code. None of the blackbox fuzzing tools can identify this constraint and maintain the required parameter dependencies across the requests. While a state-aware fuzzer is recently proposed to observe the workflow control~\cite{state-aware}, its fuzzing requests are still insensitive towards cross-request dependencies. For other approaches that consider parameter relationships~\cite{sso,shop-for-free}, they are largely manual and protocol-specific. 

%When server-side source code is available, the vulnerabilities can be mitigated by comparing the inconsistencies between client-side and server-side validations ~\cite{viewpoints}, or replicating the server-side validations to the client-side~\cite{}. They are however impractical to complex and existing applications. In practice, an implementation of banking transfer avoids processing any parameters at ServHandler$_A$ but validates the parameters only when they are con-firmed and forwarded from Step$_B$ to ServHandler$_B$. Such an implementation cannot only maintain the usability (legitimate users guided by client-side validations at Step$_A$) and security (tampered inputs bypassing Step$_A$ eventually caught at ServHandler$_B$) but also eliminate unnecessary storage or checks. Nevertheless, the proposed mitigations are to undo the optimization effort owing to the deliberate validation inconsistencies for \{Step$_A$ v.s. ServHandler$_A$\} and \{Step$_B$ v.s. ServHandler$_B$\}. More generally, given the fact that server is intrinsically more knowledgeable (e.g., access to sensitive DB) than the client, what can be validated at each side is likely inconsistent. It is difficult to efficiently annotate all expected inconsistencies especially in complex applications. 

This paper presents Cross-Request Scanner (CRS), a novel approach that respects and leverages the intended workflow and parameter dependency controls while scanning for parameter tampering vulnerabilities. CRS consists of two phases: capturing and fuzzing. In the capturing phase, CRS records a set of valid user actions, identifies the one-time tokens, tracks the cross-request parameter dependencies, and learns key features (e.g., locations of submit buttons and reflected parameters) that indicate a server acceptance from rendered server responses.

In the fuzzing phase, CRS sequentially replays the user actions to fetch new responses while keeping those confirmed one-time and dependent parameters intact so as to preserve the intended workflow and parameter dependency. Other parameters are mutated and placed back to the application itself for validations. Only those client-side rejected parameters are then forcefully submitted by bypassing client-side validations. Finally, it reports a vulnerability if the server accepts the mutated parameters and gives responses that are in line with those key features learned in the initial valid submissions (e.g., submit buttons reappeared, parameters reflected).
%In particular, we are interested in finding mutated parameters that are banned by client-side validations yet accepted by the server. 
%In the capturing phase, a human analyst provides an initial valid set of user actions by interacting with a testing application as usual in a browser. While capturing the user actions as a browser add-on, CRS tracks the parameter dependencies across requests. From the rendered server responses, it identifies key features that indicate server acceptance such as positions of submit buttons and reflected parameters. 
%Next, it replays the original set of user actions, it can thus spot out one-time tokens by isolating those parameters that differ from the initial set of requests. Those key features that are missing in the replay are also dropped. Proceeding to the fuzzing phase, it sequentially replays the user actions and keeps one-time tokens intact so as to maintain the in-tended workflow. Except for those marked dependent, parameters are mutated and put back to the application itself for validation. Only those client-side rejected ones are then forcefully submitted with the client-side validations bypassed. 

While CRS takes a set of manually provided user actions, we argue that it is unavoidable in discovering parameter tampering vulnerabilities (i.e., unlike discovering XSS by simply asserting a pop-up dialog after an injection of what comprises \verb"alert(1)"). Referring to the banking example, a parameter tampering vulnerability can be resulted by mutating only a digit in the \verb"TO" account. Therefore, it is clear that the semantic meaning and underlying consequence of a mutation can be domain-specific, and thus known only to humans. Existing work also corroborate the needs of manual effort~\cite{whyjohnny,shop-for-free}. CRS requires the least amount of manual assistance among blackbox parameter tampering scanners~\cite{notamper,acunetix,webscarab,tamperdata}.

The contributions of this paper are as follows:

\begin{itemize}

\item A field study on online banking applications to understand their workflow and implementations, and how they can be intercepted for vulnerability scanning.
\item A novel approach that respects the intended workflow and cross-request parameter dependency while fuzzing, as well as to correlate the dependency in both requests and responses for automatic detection of server acceptances.
\item An ``in-context fuzzing'' technique to build a blackbox vulnerability scanner that drives fuzzing in the application context allowing dynamic features to be preserved, and thus improving coverage and accuracy.
\item The discovery of real-world vulnerabilities that are uncovered only by our scanner, which existing approaches miss.
\end{itemize}

The rest of this paper is organized as follows: Section~\ref{sec:crs_examples} describes the online banking transfer applications of different banks. Section~\ref{sec:crs_approach} discusses the CRS approach. Section~\ref{sec:crs_implementation} provides some technical background on the web applications, followed by the implementation of the CRS scanner. Section~\ref{sec:crs_evaluation} evaluates the scanner, and details the vulnerabilities uncovered. Section~\ref{sec:crs_related} presents related work. Finally, we conclude.
\section{Motivating Examples}
\label{sec:crs_examples}

This paper focuses on real-world applications, and that makes the testing intrinsically blackbox and more challenging owing to the lack of server-side source code. This section summarizes four representative banking transfer applications, which are among the most security-critical operations being carried out over the Internet. They include Citibank, HSBC, Bank of China (BOC), and Bank of East Asia (BEA), with their headquarters respectively located at the US, UK, China, and Hong Kong. The unauthorized transfers that are made possible in HSBC and BEA are discussed in Section~\ref{sec:crs_casestudies}.

\subsection{Workflow Design}

As outlined in Figure~\ref{fig:runningexample}, all banks adopt a three-step workflow design to receive instructions of banking transfers. 

\textbf{Step$_A$}. A user specifies a source (i.e., \verb"FROM") and a destination (i.e., \verb"TO") account besides setting an amount (i.e., \verb"AMT") to transfer. The government mandates that the user can transfer his money only to third-party accounts that are authorized through an out-of-band channel. HSBC accepts unauthorized accounts to be specified in this step, and provides on-the-fly authorization in the next step, as shown in Figure~\ref{fig:hsbc}. Citibank and BOC users can pre-authorize an account beforehand in a separate online form with an One-Time Password (OTP) that is received through SMS. BEA offers only offline authorization (i.e., must register an account by person in a physical branch). Finally, the user submits Req$_A$ by clicking ``Go''.

\textbf{Step$_B$}. The user reviews the transaction details returned by the server. If an HSBC user has specified an unauthorized account in Step$_A$, he needs to enter an OTP token from his own hardware device for on-the-fly authorization, as detailed in Section~\ref{sec:crs_casestudies_hsbc}. The user finally submits again by clicking ``confirm'' to send Req$_B$.  

\textbf{Step$_C$}. The transfer instruction is acknowledged with a transaction number. The transaction details are shown for user's reference.

\begin{table*}[tb!h]
\scriptsize
\renewcommand{\arraystretch}{1.24}
\centering
\caption{The website features that are essential for effective fuzzing by blackbox web vulnerability scanners}

\begin{tabular}{ p{78pt} | c c c c c c c | c c c | c c } \hline

        & \multicolumn{7}{|c|} {\textbf{Websites Studied}} & \multicolumn{3}{|c|} {\textbf{Param. Tampering Scanners}} &  \multicolumn{2}{|c} {\textbf{Web Vulnerability Scanners}} \\
        & HSBC & BEA & BOC & HSB & Citibank & Webjet & Jetstar & CRS & \parbox[c][16pt]{30pt}{\centering NoTamper\\~\cite{notamper}} & \parbox{35pt}{\centering Proxy-based\\~\cite{acunetix,webscarab,tamperdata}} & \parbox{30pt}{\centering Traditional\\~\cite{acunetixwvs,skipfish}} & \parbox{40pt}{\centering State-aware\\~\cite{state-aware}} \\ \hline\hline
Multi-request Applications & $\surd$ & $\surd$ & $\surd$ & $\surd$ & $\surd$ & $\surd$ &  & $\surd$ & &  & $\surd$ & $\surd$  \\ 
AJAX Applications  &  &  &  &  & $\surd$ & $\surd$ & $\surd$ & $\surd$ & & $\surd$ & & \\ 
Workflow Enforcement & $\surd$ & $\surd$ & $\surd$ & $\surd$ & $\surd$ & $\surd$ &  & $\surd$ & &  &   & $\surd$  \\ 
One-time Use Tokens & $\surd$ & $\surd$ & $\surd$ & $\surd$ & $\surd$ &  &  & $\surd$ & &  &   &  \\ 
Parameter Dependency  & $\surd$ &  & $\surd$ & $\surd$ &  & $\surd$ &  &  $\surd$ & &  &   &  \\ 
Client-side Pre-processing & $\surd$ & $\surd$ & $\surd$ & $\surd$ & $\surd$ &  &  &  $\surd$ & &  &  &  \\ \hline

\end{tabular}
\vspace{-0.6em}
\label{table:crs_compare}
\end{table*}

\subsection{Workflow Implementation}
\label{sec:crs_workflow_implementation}

Notably, HSBC, BOC and BEA implement each step of the workflow in a separate page, while Citibank integrates all steps in a single AJAX page with the use of \verb"XMLHttpRequest" API. Their workflow implementations are outlined as follows:

\textbf{Step$_A$}. Client-side code is implemented to restrict the user-supplied input parameters (i.e., \verb"FROM", \verb"TO", \verb"AMT") before they can be submitted (as Req$_A$) together with a CSRF$_1$ token that is server-generated and placed in a hidden field. All banks apply some client-side pre-processing (e.g., manipulating values of hidden fields) before validating and submitting the form. The submission approaches vary from one bank to another, as detailed in Section~\ref{sec:crs_forms}.

\textbf{ServHandler$_A$}. The server verifies whether the CSRF$_1$ token matches with what was previously stored. Upon validating other parameters that represent the transaction details, the server stores them temporarily. Finally, it returns a review page in which a newly generated CSRF$_2$ token is embedded in a hidden field. Notice that besides defending against CSRF, both CSRF$_{1\&2}$ tokens can also serve the purposes of enforcing the intended workflow and preventing duplicate transactions owing to their nature of one-time uses.

\textbf{Step$_B$}. It echoes all transaction-related parameters visually for user's review, and that a confirming click will naturally include the hidden CSRF$_2$ token in the submission Req$_B$. In HSBC, an OTP parameter may be solicited for account authorization. For HSBC and BOC, this step also serves to resubmit those transaction parameters gathered from Step$_A$ by embedding them in hidden fields.

\textbf{ServHandler$_B$}. The server again verifies the CSRF$_2$ token beforehand. In HSBC, the OTP parameter, if provided, will then be verified. HSBC and BOC further ensure if some transaction-related parameters match with those stored at ServHandler$_A$. The server finally executes the transfer with the transaction parameters.

\textbf{Step$_C$}. The transaction details and a reference number are shown.

\subsection{Intrinsic Limitations of Existing Scanners}

We verify that the banks enforce the workflow design and parameter dependency across requests by empirically and systematically running a series of experiments in their banking transfer applications. We first interact with an application through its user interface (UI), and capture a pair of submission requests from a valid transaction, which is referred to as \{Req$_{A.o}$, Req$_{B.o}$\}. Next, we prepare a mutated pair of requests \{Req$_{A.m}$, Req$_{B.m}$\}, in which all \verb"AMT" parameters are incremented by ``1''.

\textbf{Neglecting Intended Workflow and One-time Tokens}. With this mutated set of requests, we however found that they were all rejected by the server through the following experiments. First of all, the requests were replayed according to a sequence of a few Req$_{A.m}$s and then some Req$_{B.m}$s, of which the sequence obviously violates the intended workflow. The requests were all rejected. We then proceed to sequentially replay \{Req$_{A.m}$, Req$_{B.m}$, Req$_{A.m}$, Req$_{B.m}$\}, but they were still rejected despite observing the workflow. 

The server actually responds differently to requests that are initiated by re-interacting with the application as usual. Therefore, when we interact with the UI and specify the same user-supplied values (including the incremented \verb"AMT") as in those mutated requests, the new requests are accepted as expected. We investigated all the requests generated through multiple times of UI interactions. We then found some distinct server-generated tokens were introduced to every request. Since replaying Req$_{A.o}$ that was once accepted by the server would result in a failure, we conclude that these tokens are of one-time uses.

\textbf{Breaking Cross-request Parameter Dependency}. As discussed in Section~\ref{sec:crs_casestudies_hsbc}, it is found that HSBC and BOC submit some transaction-related and other parameters twice in both Req$_A$ and Req$_B$. To understand the relationship between the parameters, we thus run the remaining three possible pairs of request combinations \{Req$_{A.o}$, Req$_{B.m}$\}, \{Req$_{A.m}$, Req$_{B.o}$\}, and \{Req$_{A.m}$, Req$_{B.m}$\}, in which the one-time tokens are observed. Here, all transaction-related parameters are ``slightly'' mutated one by one.

\begin{itemize}

\item When running \{Req$_{A.o}$, Req$_{B.m}$\}, it is found that some parameters mutated only at Req$_{B.m}$ are actually disregarded, while their corresponding values given at Req$_{A.o}$ are instead honored by the server. This result shows that these parameters must be mutated at Req$_A$ during fuzzing. Mutating them at Req$_{B.m}$ can be ineffective.
\item For \{Req$_{A.m}$, Req$_{B.o}$\}, server rejections are resulted since some parameter values are inconsistent between the request pair. Clearly, the server enforces certain parameters to be equal across the requests. 
\item Likewise, for \{Req$_{A.m}$, Req$_{B.m}$\}, server rejections depend on whether some parameters are mutated in such a way that the required dependency are broken across both requests.
\end{itemize}

In a nutshell, existing scanners~\cite{notamper,acunetix,webscarab,tamperdata,acunetixwvs,skipfish,state-aware} all suffer from these limitations despite the prevalence of multi-request applications. The limitations are generic and intrinsic to all web vulnerability scanners. Hence, most attack vectors in their fuzzing attempts are incapable of reaching the vulnerable code. 

These findings are confirmed by running the publicly available scanners~\cite{acunetix,tamperdata,webscarab,acunetixwvs,skipfish}, and analyzing their fuzzing patterns against a banking transfer website mimicked by us. We also carefully read through the descriptions of those tools that are unavailable to us~\cite{state-aware,notamper}. We attribute the main reason for such limitations to their fundamental scanner design. All blackbox scanners are common in crawling a website only once, and will rely on this static set of captured requests for mutation and fuzzing. Therefore, it is unsurprising that they do not even work well with applications that have applied token-based CSRF defenses. However, it is non-trivial as to how such a scanner design can be patched to identify, renew, and relate a token from a former response to a corresponding fuzzing request as well as handling the synchronization issues.

\section{The CRS Approach}
\label{sec:crs_approach}

We aim at addressing not only the problems concerned but also other dynamic characteristics that have made scanning of real-world web applications difficult. It is hard to resolve all these issues in the existing scanner design largely owing to their \textit{crawl-once-and-replay-many} and stateless fuzzing approaches.
%, not to mention other design considerations to be discussed in Section~\ref{sec:crs_considerations}. 

Hence, this paper proposes a parameter tampering scanner called Cross-Request Scanner (CRS), which structurally changes the traditional fuzzing approach. Table~\ref{table:crs_compare} provides a quick comparison of various scanners, which shows their support to those website features that are essential for effective fuzzing. Here we outline our approach and its design considerations, which has enabled the discovery of some previously unknown vulnerabilities.

\begin{figure}[htp]
\centering
\includegraphics[width=0.48\textwidth]{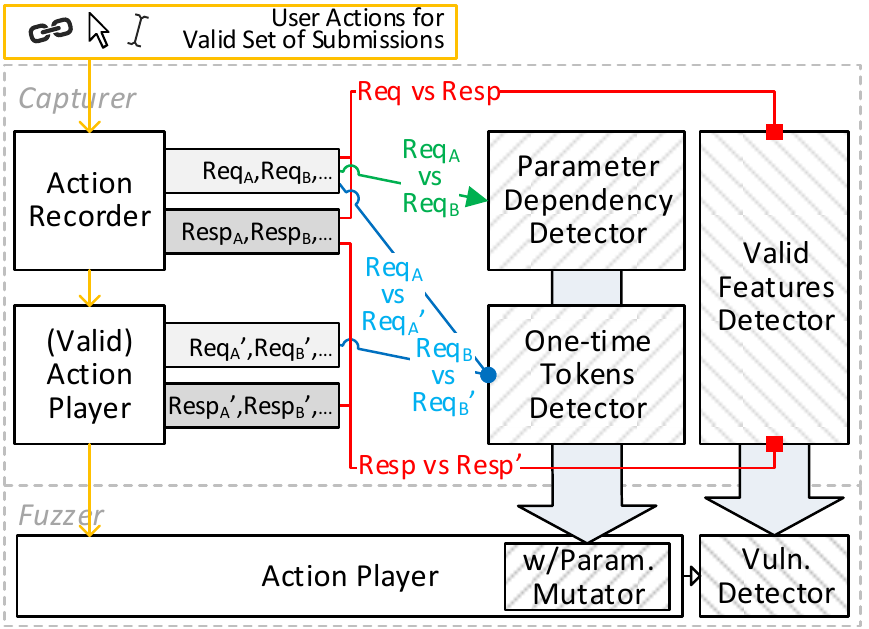}
\caption{The architecture of Cross-Request Scanner (CRS)}
\label{fig:arch}
\vspace{-1em}
\end{figure}

\subsection{Overview of CRS}

Figure~\ref{fig:arch} depicts the high-level architecture of CRS, which is a blackbox parameter tampering scanner designed to find parameters that are client-side rejected yet server-side accepted~\cite{immutable}. It comprises the capturing and fuzzing phases:

\textbf{Capturing Phase}. While running CRS as a browser add-on, an analyst provides an initial valid set of submissions by interacting with a testing application as usual in the browser. CRS captures the user actions for later replays. Meanwhile, it tracks the parameter dependency across the submission requests, and thus a parameter with its value being the same as in the last request is marked dependent. From the server responses, it identifies some candidates of key features such as locations of submit buttons and reflected parameters, that can represent an acceptance by the server. It then replays the original user actions to get a second valid set of submissions, and classifies those parameter values that differ from the initial submissions as token candidates. In these two valid submissions, those key feature candidates that cannot be reproduced will be dropped. Finally, it confirms that each token candidate is of one-time use if a rejection is resulted upon repeating the same request. 

\textbf{Fuzzing Phase}. To observe the intended workflow and one-time use tokens, CRS sequentially replays the user actions and keeps the tokens intact during fuzzing. Each parameter is mutated and placed back to the application to undergo its client-side validations. The dependent parameters are exempted from fuzzing after the candidates are mutated, submitted, and actually confirmed to result in server rejections. Next, those parameters that are rejected by the client-side are forcefully submitted with the validations bypassed. With such a rejected parameter, CRS reports a vulnerability if the server reacts in such a way that the expected workflow and key features can be reproduced as in the valid submissions.

\subsection{Design Considerations}
\label{sec:crs_considerations}
Here we explain the underlying considerations of the approach.

\textbf{Preserving Intended Workflow}. Sequentially fuzzing a whole set of requests, one after another, is a necessity to preserve the intended workflow. CRS achieves this by replaying user actions.

\textbf{Respecting One-time Use Tokens}. One-time use tokens are by definition those server-generated parameters that differ across two requests even though an identical set of user actions are provided. Before any user inputs are entered, these tokens are readily present either in hidden fields or the query string of submission URLs. This important clue can help eliminate many false classifications. However, proxy-based scanners~\cite{acunetix,webscarab,tamperdata} which capture a request at proxy level (i.e., outside the browser context) obviously possess no such knowledge. To extract these tokens, complex state maintenance and even synchronization issues are anticipated if we are to modify existing scanners to fetch a previous response for token extraction before making every fuzzing request. In contrast, CRS proposes to identify and fuzz directly in the browser so that these tokens are inherently recognizable and well-preserved. After confirming that they are of one-time uses, CRS excludes them from being mutated during fuzzing.

\textbf{Respecting Parameter Dependency}. Given the two valid sets of requests, CRS marks a parameter as dependent candidate if its value is equal to any parameters in the last request. This simple classification might be an overkill, so CRS still fuzzes each dependent candidate once for confirmation. Hence, only those that result in server rejections will be confirmed as dependent. To preserve the confirmed dependency, the fuzzer mutates such a parameter only in the first request it appeared but not in any successive ones. For instance, if a parameter value in Req$_B$ is confirmed to be dependent to one in Req$_A$, then it is mutated only at Req$_A$ but kept intact at Req$_B$. In fact, this algorithm is also good for locating session-based tokens of which the values are present in multiple requests and will expire only after the session is terminated.

\textbf{Preserving Client-side Preprocessing}. Web applications implement their client-side logic using JavaScript. Without any page loading, modern applications may dynamically add or drop input fields in response to a user action. In the banking examples studied, HSBC, BOC, HSB and BEA even manipulate some hidden fields before performing form validations and submissions. AJAX-driven websites such as Citibank use the asynchronous \verb"XMLHttpRequest" API. Existing scanners are however ignorant to all of them. It may look intuitive to preserve these features simply by running the applications in a full-blown browser during testing. But this actually poses a challenge in preserving these client-side executions while mutating parameters, to be detailed in Section~\ref{sec:crs_vul_detector}.

\textbf{Detecting Client-side Rejection}. Some existing approaches infer the validation rules by coding analysis\cite{notamper,waptec}. Given a blackbox environment, it is however difficult to preserve complex and dynamic validations that may even depend on AJAX feedbacks. Instead, to be generic, CRS places a mutated parameter back into the application itself to determine whether a submission is forbidden, and if so, the parameter must be rejected by client-side validations and will be used for fuzzing. Concerning the mutation algorithm, it heuristically derives new values (e.g., $+1$, $\times-1$) from either a user-given or default value besides adopting some static values from~\cite{tamperdata}. Although it may not be as exhaustive as what can be generated through code analysis, it suffices in the current study.

\textbf{Detecting Server-side Acceptance}. Most existing blackbox scanners use edit distances for response classifications~\cite{notamper,webscarab,acunetix}. However, this approach often fails in classifying JSON-formatted~\cite{json} feedbacks that are commonly used in AJAX applications. For responses such as \verb"{success:1}" and \verb"{success:0}", the zero edit distance between them makes server acceptances hardly discernible from rejections. In particular, NoTamper~\cite{notamper} also requires an analyst to provide a baseline pair of valid and invalid submissions, such that a fuzzed response can be clustered to the one with closer edit distance. In contrast, CRS requires only one valid set of submissions, and is able to figure out the invalid responses which must fail to reproduce the expected submit buttons, workflow, and reflected parameters. The concept of button detection is similar to a proposal that clusters server responses into different states according to their outgoing hyperlinks~\cite{state-aware}. But here, CRS can support those responses that may contain no hyperlinks and those with indistinguishable edit distances. It is because the classification is applied to those DOM elements that receive visual changes~\cite{mutationobserver} upon AJAX feedbacks, rather than tapping directly to the response data.

\section{Implementation}
\label{sec:crs_implementation}
To realize our approach as a usable tool, this section first presents a background study on how parameters are gathered, validated, and submitted in web applications in general. Next, we detail the implementation of the core components, including the Capturer and Fuzzer, of the scanner.

\subsection{Technical Background of Web Applications}
\label{sec:crs_techbackground}

\subsubsection{Sources of Parameters and Validations}

An HTML form typically encloses a variety of input controls to take users' input, and each of which is associated with a name and optionally a default value (e.g., with the \verb"value" attribute). Such a name/value pair then contributes a request parameter during form submission. On the other hand, server-generated values are typically embedded in hidden fields, or hardcoded at the query string of submission URLs.

The choice of an input control can obviously impose a restriction on the input format that a user is supposed to follow. For instances, two radio boxes denoting genders with values defaulted to \verb"M" and \verb"F" can prevent users from entering unexpected values, whereas the use of a hidden field is a means of validation that it must be equal to the given (or default) value.

In addition, applications can be programmed to apply explicit client-side validations with JavaScript and HTML 5 API~\cite{html5}. For instances, application typically enforces string-valued input to be composed of only alphanumeric characters, and that numeric input to be within a proper range (e.g., $age >= 18$). If a form fails the validations, the submission must be forbidden. 

These explicit validations and those imposed by input controls are all referred to as client-side validations. Notice that they are implemented only for improved usability. Input validations must also be implemented at the server for security purposes.

\subsubsection{Form Submissions}

\label{sec:crs_forms}

Client-side validations must be applied before an actual submission by registering an event handler to either the \verb"submit" event of the form or the \verb"click" event of a clickable element. In this paper, the former approach is referred to as the \textit{event-based form submission}, and the latter one as \textit{programmatic form submission}, whereas those submitted using the \verb"XmlHttpRequest" API are referred to as the \textit{AJAX form submission}. Their behaviors, as detailed below, essentially govern how they can be intercepted for fuzzing:

% \begin{itemize}
\textbf{Event-based Form Submission}. When a user clicks a submit button (e.g., \verb"<input" \verb"type=image|submit>"), or hits ``Enter'' in a text field, the \verb"submit" handler of the form is invoked. If its return value is set to \verb"true", the form proceeds to submit and will trigger a page load; otherwise, the submission can be forbidden. JavaScript validations are applied inside this handler, and thus the form is submitted only if it is properly validated (i.e., \verb"return validated(form)").

\textbf{Programmatic Form Submission}. When employing an (click) event that is non-specific to form submission, the event handler must explicitly invoke the \verb"submit()" method for a form submission. Clearly, this approach will trigger a page load but not the \verb"submit" event. The invocation is typically restricted by the form validity (i.e., \verb"validated(form)" \verb"&&" \verb"form.submit()").

\textbf{AJAX Form Submission}. Modern applications submit a form at the background without reloading the page. To achieve this with the \verb"submit" event handler, \verb"false" is always returned to cancel the default submission and thus the page load. There is no page load problem when using other event handlers such as \verb"onclick". When the form passes the validations, the event handler will invoke a custom function (e.g., \verb"sendAJAX()"), that is made to encode the name/value input pairs as request parameters, and finally submit a request using the well-known \verb"XMLHttpRequest" API.
% \end{itemize}

% \begin{figure}[!t]
% \centering
% \includegraphics[width=0.45\textwidth,trim=0 17 0 15]{fig/table2.pdf}
% \caption{(a) JavaScript validations executes in a handler registered to the (b) submit event of the form or (c) click event of an element.}
% \label{fig:code}
% \end{figure}

\subsection{The Capturer of CRS}
When the add-on is being launched alongside the browser, the Capturer actually begins by recognizing the form submission approach that is being used. Even before any user actions, it has already interposed on the \verb"submit()" method of every form, and registered an handler to the \verb"beforeunload" event. If the \verb"submit()" method is called, the form submission approach is a programmatic one. If the \verb"beforeunload" handler is first fired (i.e., not initiated by the \verb"submit()" method), then it is clearly an event-based form submission. Otherwise, the form submission is using AJAX.

Here we present the implementation of the smaller modules including the Action Recorder, Action Player, Parameter Dependency Detector, One-time Token Detector, and Valid Features Detector.

\subsubsection{Action Recorder}
The Action Recorder is responsible for capturing the URLs, HTTP requests and responses of an initial valid set of submissions. The request URL and parameters are obtained from the attributes and input elements of the corresponding form, while the responses are also readily accessible by similar DOM API calls.

On the other hand, it listens to all keystrokes and clicking events, and takes the unique id attributes or otherwise XPath positions~\cite{xpath} as references to those target elements that receive the events. This is implemented by registering event handlers at the \verb"document" level, so that they are recorded before propagating down to the target elements or possibly canceled by the intermediate ones. 

\subsubsection{Action Player}

With the captured user actions, the Action Player of the Capturer can replay them to obtain the second set of valid submissions. This is achieved by synthesizing and replaying the events with the \verb"initEvent()" and \verb"initMouseEvent()" APIs. In general, the implementations of the event recorder and player is similar to Selenium, which is an open-source recording and replaying tool~\cite{selenium}.

\subsubsection{One-time Tokens Detector}
This detector concerns only server-generated parameters that are originated from hidden fields and query string of submission URLs. Working as a browser add-on, CRS can easily sort out these parameters by DOM inspection. The name/value pairs in the query string are then properly decoded using the \verb"decodeURIComponent()" API. Among these server-generated parameters, it identifies those parameters that have the same name yet distinct values across each corresponding pair of requests between the initial and second valid sets of submissions. 

Each of these identified candidates is confirmed of one-time use if making an identical request will indeed lead to a server rejection. Given a non-AJAX form, it is relatively easy to tamper the corresponding hidden field or submission URL and reuse the token in order to spoof an identical request. A server rejection, which is literally a lack of any of the key features, can then be asserted. However, for forms submitted over AJAX, more complex techniques to be detailed in Section~\ref{sec:crs_vul_detector} and~\ref{sec:crs_valid_features} are similarly used here to spoof an identical request as well as to find if any key features are missing. The names of these confirmed parameters are recorded, and they will be preserved when fuzzing other parameters. 

\subsubsection{Parameter Dependency Detector}
This detector is actually responsible only for identifying candidates of dependent parameters, which are subject to further confirmation by the Fuzzer. First, CRS applies regular expressions to determine the value types. For instance, \verb"/^[\d,]+(?:\.\d+)?$/" is used to test for numbers. With any commas stripped, the numeric values are casted into float using the \verb"parseFloat()" API, while all other values are trimmed. When comparing the parameters across the requests of the initial valid set of submissions, it neglects the names but considers only those transformed values. A candidate parameter is finally considered dependent to another parameter of the last request only if two values are identical. The references (i.e., id attribute or XPath location) to these candidates are recorded.

\subsubsection{Valid Features Detector}
\label{sec:crs_valid_features}
This module is to determine the key features that indicate reproducible acceptances by the server. The references to the candidate elements of key features are recorded throughout the initial valid set of submissions. In the second valid set of submissions, the extracted features are relocated in the rendered server responses. Hence, the final key features to be used are all verified to be reproducible in both valid set of submissions. 

First of all, the buttons that are clicked and captured in the Action Recorder automatically qualify as the key features since they are always needed to follow the intended workflow. 

Next, CRS extracts a number of other key features by locating user-supplied parameters that are visually reflected in the rendered responses. This is achieved by searching for the presence of a parameter value in the \verb"textContent" and \verb"value" attributes of every leaf node through DOM tree traversal. This technique is commonly used in search engine keyword highlighting~\cite{keyword-highlight}. Nevertheless, CRS handles number matching differently in order to tolerate formatting issues that are introduced by the server. For instance, \verb"12345" is first converted to a regular expression object, i.e., \verb"\/1[^\d]?2[^\d]?3[^\d]?4[^\d]?5\/", and can thus find its presence in a node content such as \verb"$12,345.00". If a value has resulted in multiple occurrences that exceed a configurable limit\footnote{For instance, a single-character value such as 1 or 0 will likely occur everywhere. While a maximum of 3 times is generally reasonable for applications that reflect parameters, here we do not quest for a magic number but leave it configurable.}, the candidate parameter is disqualified. Obviously, each parameter is mutated based on its specific features.

In case no parameter reflections can be identified, CRS will heuristically search for some keywords that indicate sever acceptance with the same DOM traversal approach. The case insensitive keyword choices include ``success, done, complete(d), execute(d), ok(ay), or update(d)'' but must also exclude ``not, n't, fail(ed), err(or) or sorry''. Since the extension to multi-lingual support is trivial, so a string is now tested only as follows:

\begin{lstlisting}[language=JavaScript,numbers=none]
/\b(?:success(?:ful)?|done|completed?|executed?|ok(?:ay)?|updated?)\b/i.test(string) 
&& !/\b(?:not|sorry|fail(?:ed)?|err(?:or)?)\b|(?:[a-z]n\'t\b)/i.test(string)
\end{lstlisting}
\vspace{+18pt}

\setlength{\parindent}{0in}The approximate string matching algorithms~\cite{approx-string} may better perform in some applications, but they are not used here for less complexity. In case of no identifiable features, the analyst will be prompted to provide a regular expression statement for asserting the server acceptance. We so far did not encounter such a manual need in our experiments.

\setlength{\parindent}{0.2in}The CRS approach focuses on the rendered responses to locate key features even for AJAX applications. It generally ignores the raw response data as replied from the server, but simply allows the testing application itself to react accordingly and make DOM changes. In this regard, CRS is so designed to leverage the application itself to generate the required \verb"XMLHttpRequest"s. So, the relevant \verb"onstatechange" handlers that respond to the AJAX feedbacks can be triggered.

On the other hand, CRS limits the scope of feature searching to accurately the part that changes after the AJAX form submission. To implement this, CRS registers a callback function to the \verb"MutationObserver" API~\cite{mutationobserver} right before the submit button is clicked. After the application has made some DOM changes, the callback function will be invoked with those DOM objects that receive the changes, where the feature extraction algorithm will be applied to those visible objects. Only if no visual changes are ever made, CRS will use the HTTP status code and response data to determine server acceptances. However, we did not encounter such a case in the websites we tested.

\subsection{The Fuzzer of CRS}
The Fuzzer begins by confirming the list of dependent parameters. It replays the captured user actions up to the page where a candidate dependent parameter is found. The value of which is then mutated to ensure that it differs from that of the last request to make a dependency breaking attempt. Finally, the dependency is confirmed if a server rejection is encountered. To assert this, a fuzzing technique that is also applied to other non-candidate parameters is used and described as follows.

\begin{lstlisting}[float=left,xleftmargin=2em,language=JavaScript,label=list:fuzzer,caption=The core algorithm of the Fuzzer]
while (action = userActionList.next()) {
	if (action.isSubmissionClick() 
			&& param = paramList.nextCandidate() 
			&& !param.isOneTimeToken() 
			&& !param.isDependencyDetected()) {
		getClientRejected(param)
		form.forceSubmission()
		detectServerAcceptance()
	} else {
		action.replay()
		paramList.addNewlyAppearedFields()
	}
}
function getClientRejected(param) {
	if (mutated = param.nextMutated()) {
		form.interceptSubmission(function() {
			form.cancelSubmission()
			getClientRejected(param)
		})
		form.fillIn(mutated)
		action.replay()
		paramList.addNewlyAppearedFields()
	}
}
\end{lstlisting}

Listing~\ref{list:fuzzer} outlines the core algorithm of the Fuzzer. It sequentially replays the captured user actions (Line 1 \& 10) and stops before a submission click (Line 2). Parameters that are confirmed of one-time use (Line 4) or dependent to a previous request (Line 5) are exempted from fuzzing. Other parameters are mutated one by one with respect to its default or user-given value (Line 15). The mutated parameter is then placed back to the form, and the corresponding events (e.g., \verb"onclick", \verb"onchange") are also triggered (Line 20). Any newly appeared fields will also be captured here for later mutation (Line 11 \& 22). Upon replaying the submission click event (Line 21), the web application itself validates the form concerned. If a submission is attempted, that means the mutated parameter is accepted by the client-side validations (Line 16). The Fuzzer will however cancel the submission (Line 17), and repeatedly mutate the parameter until it is rejected by the client-side (Line 18). With that client-side rejected parameter, the Fuzzer bypasses the validations to force a form submission (Line 7). Finally, it asserts a server acceptance by locating all the valid key features (Line 8), and is able to report those parameters that are rejected by the client-side validation but accepted by the server as vulnerable to parameter tampering attacks.

Those smaller modules including the Parameter Mutator and Vulnerability Detector are detailed as follows. The discussion on Action Player is omitted as it is the same as that in the Capturer.

\subsubsection{Parameter Mutator}
This module is responsible for generating mutated values. The default values, or otherwise analyst-provided ones form the bases for mutations. First, CRS mutates the numeric portion of this existing value in order to derive some integral, decimal, negated, incremented, decremented, and multiplied versions (e.g., \verb"parseInt()", $+n/1000$, $-n$, $+n$, $\times -1$, $\times n$, where $n$ comes from some hardcoded and randomized integers). 

CRS then determines the input types by inferring from the whole value (e.g., \verb"/[\d,\.]+/" for numbers), type attributes (e.g., \verb"date" from HTML 5), class attributes (e.g., alphanumeric, datepicker; as often employed by JavaScript validation libraries~\cite{jquery}), name attributes, and field labels (e.g., tel, date). For boolean values that are likely represented in \verb"1", \verb"0" ,\verb"Y", \verb"N", \verb"T", \verb"F", \verb"true", or \verb"false", new values are derived by negating the boolean. For certain input types such as date, time and percentage, CRS adds some relevant hardcoded values (e.g., \verb"2013-9-22", \verb"23:59", $111\%$). Optionally, the Mutator is configurable to also output test cases for violating the input types and other field restrictions such as \verb"required" and \verb"maxlength" (e.g., concatenating an existing value with itself by multiple times). 

Here some of the hardcoding values are borrowed from a tool called TamperData~\cite{tamperdata}. It is certain that this heuristic input generation module is imperfect, and can be further enhanced by performing extra coding analysis~\cite{notamper,waptec,viewpoints}. However, the current setting is found sufficient in our experiments.

\subsubsection{Vulnerability Detector}
\label{sec:crs_vul_detector}
This subsection mainly enumerates how to intercept and cancel submission attempts (client-side accepted parameters) as well as force a submission (client-side rejected parameters) for each form submission approach. Interestingly, this involves flipping the original decision on whether to proceed a submission. With a fuzzing attempt that is client-side rejected, CRS can then discern server acceptances by detecting those valid features found by the Capturer.

\textbf{Intervening in Event-based Form Submission}. To intercept the submission, CRS interposes on the \verb"onsubmit" handler of the form to wrap the original one.
%\footnote{We omit here how to interpose on the submit handlers that are registered with a DOM level 2 event model.}. 
Our handler disables HTML 5 automatic validations by setting \verb"form.noValidate" to \verb"false". It then discerns the parameter validity by examining the return values of the original \verb"onsubmit" handler and \verb"form.checkValidity()" method that respectively reflect the results of JavaScript and HTML 5 validations. If an accepted parameter is encountered (i.e., both returned \verb"true"), CRS will return \verb"false" in our \verb"onsubmit" handler to cancel the submission. For rejected parameters, CRS forces the submission by returning \verb"true" in our handler.

\textbf{Intervening in Programmatic Form Submission}. To intercept the programmatic submission, CPS overrides the \verb"submit()" method of the form. To discern parameter validity, we cannot rely on the return value which is always \verb"void". Instead, we know that our method will be called only with client-side accepted parameters. We will then simply ignore the submissions for these parameters. For other (i.e., rejected) parameters, CPS deliberately calls the original \verb"submit()" method to force submissions. 

\textbf{Intervening in AJAX Form Submission}. Intervening in AJAX submissions is the most challenging implementation for the following reasons. First, we do not assume web applications to use any particular JavaScript libraries such as jQuery~\cite{jquery}. Second, in contrast to the previous approaches of which the event and method are directly associated with a form object, \verb"XMLHttpRequest" multiplexes all kinds of HTTP requests that are non-specific to form submissions (i.e., without a function like \verb"form.ajaxSubmit()" that can be intercepted similarly). Third, owing to the asynchronous nature, multiple AJAX requests can simultaneously exist, such as refreshing a placeholder for advertisements and dynamic validations that involve server feedbacks (e.g. nickname uniqueness). Hence, it is non-trivial to distinguish which request is responsible for the form submission concerned.

We solve this problem by first intercepting the methods \verb"open()" and \verb"send()" at the prototype level of \verb"XMLHttpRequest", which are invoked before making a request. It is necessary to intercept both \verb"open()" and \verb"send()" methods, through which input parameters are provided respectively for GET and POST requests.

To distinguish the particular request that corresponds to the form submission, CPS appends a nonce to the value of an existing (hidden) input control\footnote{Alternatively, we can inject a hidden field with a nonce value, so that it will be submitted along with other field values.}, of which its validity is unaffected. Hence, the AJAX request that consists of this nonce can be isolated in our intercepted methods. Here, an invocation of our intercepted \verb"send()" method implies that the parameter is client-side accepted. CPS discards the submission by ignoring it. 

For client-side rejected parameters, the application will not even instantiate the \verb"XMLHttpRequest". Given a simple application, it may be feasible to reproduce the instantiation by program slicing. But we instead develop a more succinct approach. CPS achieves a forced submission by reusing the set of valid parameters to trigger a submission request. But then the parameters are substituted with the client-side rejected ones at the intercepted methods. The nonce is also removed during so before sending the requests.

\section{Evaluations}
\label{sec:crs_evaluation}

This section focuses on evaluating the proposed scanning approach in terms of its practicality and effectiveness. Practicality considers how well CRS can automate and intercept real-world applications, while effectiveness measures the vulnerability scanning capabilities. All experiments are conducted with an entry-level PC using Firefox. The lightweight implementation actually incurs only negligible performance overhead, and that the scanning time is dominated by the server responsiveness and a configurable delay deliberately introduced to avoid ``DoS-ing'' the servers. We thus omit here a benchmark on performance overhead. If needed, the fuzzing time can be further reduced by using a headless browser~\cite{phantomjs}.

This section also briefly discusses some insights behind the severe vulnerabilities uncovered with the scanner, and shares how they are responsibly reported and subsequently resolved~\cite{shop-for-free}.

\subsection{Ethical Concerns}
For the practicality evaluation which generates no tampered requests, there are no ethical issues. For the effectiveness evaluation, we turned tampering on in real-world web applications. In the worst case scenarios, the tampered banking transfer requests could possibly result in money transferred either from our account to an unknown account or from an unknown account to ours. We thus tested only the interbank transfer applications which all require at least two days to take effect, and hence we have sufficient time to manually cancel (or make instant transfer before) any unwanted transactions. To further limit the risk, all requests are capped to use only a small amount of money during the automated fuzzing. To confirm a vulnerability to be a working exploit, we only manually made online transfers from and to accounts that are owned by us at different banks. All findings were promptly disclosed to the affected parties. We were grateful for being formally recognized by the government and banks' officials as ``good citizens''.

\subsection{Empirical Results}
We tested a total of seven applications, including five online banking transfer applications including HSBC, BEA, Citibank, BOC, and Hang Seng Bank (HSB), followed by two traveling websites including Jetstar airline and an online travel agent Webjet. These banks are chosen since their branches are in close geographical proximity to our institution, so that we can minimize the overhead of visiting them for opening bank accounts. The traveling websites are chosen because of a plan to visit Australia, where the headquarters of Jetstar and Webjet are located.

\textbf{Practicality Evaluation}. We found that CRS can practically drive all the testing applications to run successfully. In this regard, CRS outperforms the existing scanners, which fail to support many critical features that are required for effective scanning, as tabulated in Table~\ref{table:crs_compare}. Some scanners cannot even run the applications due to their lack of JavaScript and AJAX support. 

We are unaware of any unsupported form designs except those with non-standard input controls and those built with plug-ins such as Adobe Flash. This limitation is however non-specific to CRS but also other parameter tampering scanners~\cite{notamper,waptec,acunetix,tamperdata}.

\textbf{Effectiveness Evaluation}. We evaluated that CRS is effective in uncovering some previously unknown vulnerabilities even in banking websites. Table~\ref{table:data} summarizes the confusion matrix of the vulnerability scanning results, which is further discussed below.

\begin{itemize}

\item\textit{True Positives}. The vulnerabilities uncovered in two banking websites are separately detailed in Section~\ref{sec:crs_casestudies}. CRS discovered another vulnerability in an AJAX page of the official Jetstar website. There is a restriction on tracking on-time performance only from the last 10 days, but it is found bypass-able. We confirmed manually that it allowed us to check the performance of a randomly selected flight, which departed even three years ago. Given that the information was once (when within last 10 days) publicly accessible, we did not approach the organization for a patch. This finding serves to demonstrate the support of AJAX applications by CRS.

\item\textit{True Negatives}. In our experiments, CRS found no vulnerabilities in BOC, Citibank and HSB. We manually analyzed their client-side source code and classified them as true negatives largely because the servers have mapped each account number to a unique index (rather than using the actual account number). Thus, a tampered index could not go beyond a pre-defined mapping of authorized accounts that is maintained at the server. It is indeed a proper parameter tampering defense, as recommended by OWASP~\cite{owasp-insecure-direct-object}.

\item\textit{False Positives}. We are unaware of any false positives so far.

\item\textit{False Negatives}. Finding false negatives requires a set of known and unpatched vulnerabilities, which is intrinsically difficult in our blackbox evaluation. We managed to manually discover only one vulnerability in a travel booking website Webjet. It was missed because the lack of domain knowledge on airport code. More specifically, only some Australia airports such as HBA would cause the vulnerability. This limitation is non-specific to CRS but also any other scanners. We believe that this vulnerability is discoverable only by chance even with a human (when planned to visit Australia). We contacted Webjet, which appreciated our finding and fixed the vulnerability.

\end{itemize}

\begin{table}[t]
\centering
\renewcommand{\arraystretch}{1.2}
\caption{The confusion matrix of the CRS scanning results}
\small
%\footnotesize
\begin{tabular}{l | c c } \hline 
	& \textbf{Positives}		& \textbf{Negatives} \\ \hline
\textbf{True}   & HSBC, BEA, Jetstar 	& Citibank, BOC, HSB  \\
\textbf{False}  & /                     & Webjet \\\hline
\end{tabular}
\label{table:data}
\vspace{-0.5em}
\end{table}

\begin{figure*}[tbp]
\centering
\includegraphics[width=0.95\textwidth]{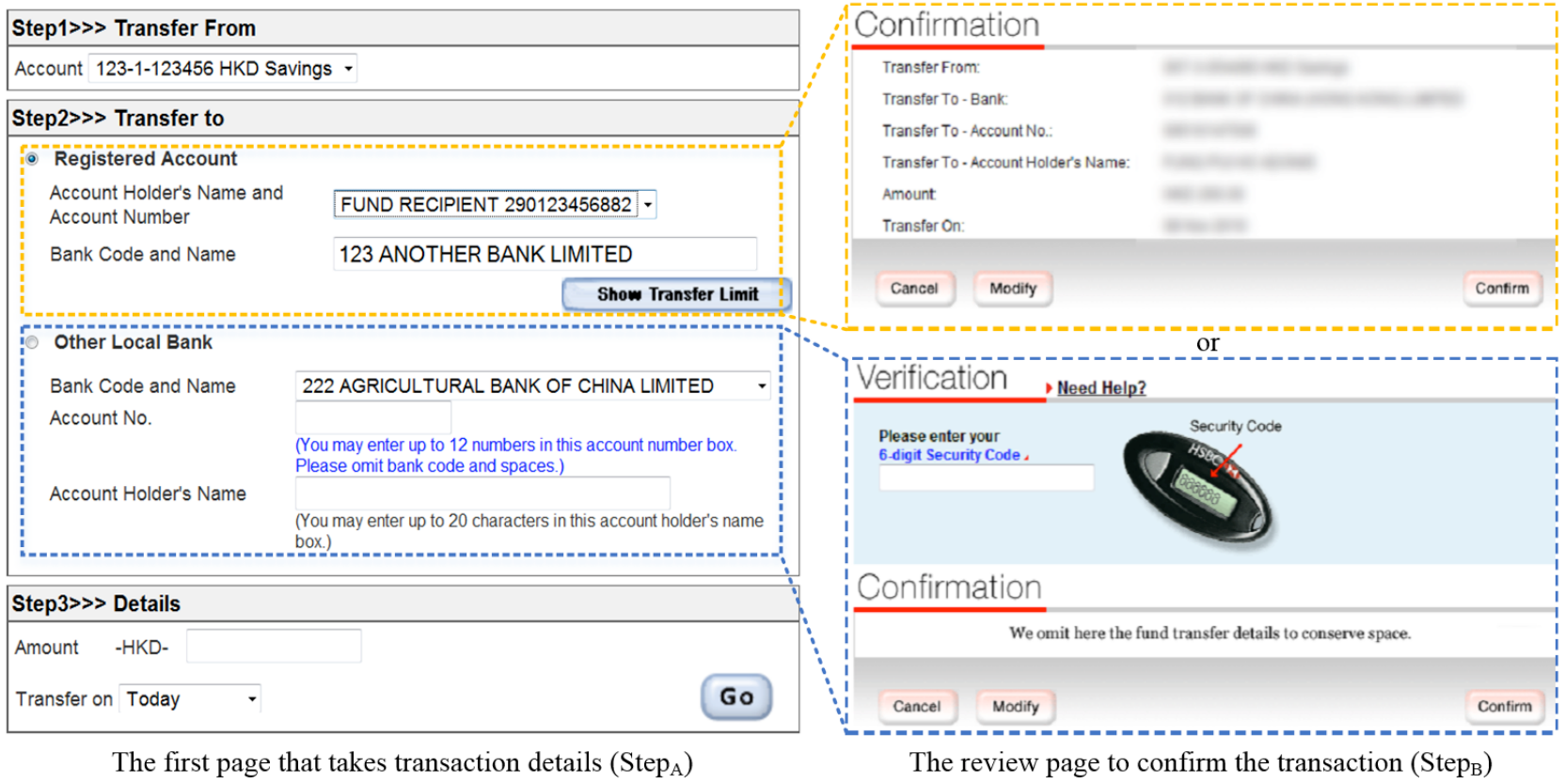}
\caption{The online banking transfer application of HSBC responds with a different review page based on the choice of radio buttons}
\label{fig:hsbc}
\vspace{-0.5em}
\end{figure*}

\subsection{Case Studies and Vulnerability Disclosure}
\label{sec:crs_casestudies}

Here we discuss the parameter tampering vulnerabilities uncovered in HSBC and BEA by CRS, as well as our encounters of responsible vulnerability disclosures. We once expected banks to be well scrutinized by existing scanners and security audit. For instance, the Acunetix web vulnerability scanner being used by HSBC is also adopted by NASA and even the Pentagon~\cite{acunetixwvs}. Being one of the largest multinational banks, HSBC also won the ``Best Information Security Initiatives'' award in the world's best consumer and corporate Internet banks~\cite{hsbcaward}. However, their existing mitigation approaches have apparently failed. Their application designs may provide some insights on why the vulnerabilities would evade the detection of their existing approaches.

\subsubsection{Bypassed OTP Requirement in HSBC}
\label{sec:crs_casestudies_hsbc}

Figure~\ref{fig:hsbc} shows two screenshots of the interbank transfer application of HSBC, which correspond to Step$_A$ and Step$_B$ in Figure~\ref{fig:runningexample}.

\textbf{Legitimate Use Cases}. When transitioning from Step$_A$ to Step$_B$, the server first verifies the one-time token and cross-request parameter dependency. The logic flow then depends on a radio button that indicates whether the account \verb"TO" is authorized (registered).

If the \textit{upper} radio button is chosen, the server will make use of the corresponding ``Account Holder's Name and Account Number'' and ``Bank Code and Name'' fields for further format verification. Therefore, the account number format should match with a regular expression \verb"/^[a-zA-Z]{1,20}~~\d{1,12}$/". It is however unlikely that checksum verification can be applied to the \verb"TO" account number since its definition is totally up to another bank. Given an authorized account, the user still has to click ``confirm'' in the review page. 

If the \textit{lower} radio button is chosen, the server will take an authorized account from the corresponding ``Bank Code and Name'', ``Account No.'', and ``Account Holder's Name'' fields. The review page will then ask the user for an OTP generated from a dedicated hardware device. This second factor authentication is to assure that the transfer is authorized by the legitimate user.
In either case, the bank sends out an SMS to the user. The \verb"TO" account number is however partially masked, and that neither the holder name nor registered status are provided.

\textbf{Vulnerability Details}. CRS is first provided with an initial valid set of user actions that go through the complete workflow to transfer \$1 from our account to an authorized account. CRS then automates the rest of the discovery, and reported several parameter tampering possibilities for the  \verb"TO" parameter. Among the scanning results, the simplest possibility was to increment the numeric portion of the original value by one (i.e., \verb"FUND RECEIPIENT~~290" \verb"123456883"). 

We manually confirmed this vulnerability by transferring money out from our HSBC account, and were able to receive it in an unauthorized account of our own in another bank. It is likely that the mutated (i.e., unauthorized) account number was handled by the logic branch responsible for authorized accounts. Hence, the most-trusted OTP was completely bypassed! It is reasonable that even a transaction-signing device~\cite{hsbcnewdevice} could not help in this regard. 

As a consequence, an attacker transferring his money to an ``unauthorized'' account can dispute the transaction on the basis that it does not correspond to a use of OTP. Alternatively, he can steal money from 
%some victims if he can hijack their sessions or simply purchase credentials from underground communities.
those victims whose sessions can be hijacked or whose credentials can be purchased from underground communities.

Existing scanners miss this vulnerability due to the enforcements of one-time use tokens, cross-request parameter dependency, and client-side pre-processing.

\textbf{Disclosure}. It was threatening that two out of four banks (by the time we ran the first set of experiments) were confirmed as being vulnerable, in spite of the limited applications we scanned. This made us believe that the prevalence of parameter tampering vulnerabilities were heavily under-estimated in online banks. Having failed to contact the right representative of HSBC, we anonymized its name and reported our findings to the government authority responsible for maintaining monetary and banking stability. The authority took our findings seriously and met with us in person to avoid any discloses through a wire. Promptly after the meeting, the authority alerted all licensed banks to request them for a check and possible fix, which may explain that no further vulnerabilities were later found in other banks. It later forwarded to us a technical clarification request from HSBC, which rolled out a patch in two days. The authority and HSBC expressed their gratitude in writing.

\subsubsection{Unauthorized Transfer in Bank of East Asia}

Here we outline only the findings that are specific to BEA.

\textbf{Legitimate Use Case}. The interbank transfer application of BEA allows transfers only to registered accounts, which can be authorized only by visiting a physical branch. In other words, one can never make an online transfer to any unregistered accounts as a protection. No SMS and email notifications were implemented. 

\textbf{Vulnerability Details}. Similar to HSBC, CRS was able to report a vulnerability that result in unauthorized transfers owing to the lack of proper server-side validations. We manually confirmed the vulnerability by transferring \$10,000 from our account to an unregistered account of our own.

We noticed an interesting implementation that a hidden field called \verb"MACcode" is set at the client-side before any form validations and submission. Its value is prepared by concatenating the transaction parameters (i.e., \verb"FROM", \verb"TO", \verb"AMT") and encoding the result by the Base64 scheme. The server was found rejecting those requests if the \verb"MACcode" received does not align with the transaction parameters. 
Since CRS places a mutated parameter back to the application, the parameter can undergo the same \verb"MACcode" algorithm at the client-side. So, it is clear that such an algorithm cannot protect message authenticity, as opposed to its confusing name.

Existing scanners miss this vulnerability owing to the enforcements of client-side pre-processing and intended workflow.

\textbf{Disclosure}. More than a week after our experiments, we disclosed the vulnerability directly to BEA. Apparently, no monitoring alarms were triggered regardless of the large transaction amount. They acknowledged our finding, and promptly fixed it within a day. Their system was later enhanced by dispatching email notifications.

% \subsection{Insight}
% It is common in the development cycle that a web designer is responsible for first producing mock-up designs (which are often images without functional features), and then the finalized design is given to an engineer to implement the underlying features. Given such a visual form design as depicted in Figure~\ref{fig:hsbc}, it is indeed intuitive for the engineer to separate two procedures in handling registered and unregistered transfers. It is even more natural to think that the OTP validation logic belongs only to the branch (e.g., \verb"else if(!reg)") that handles input fields of unregistered accounts. However, this is obviously incorrect given the vulnerability discovered. This pattern does not seem to be only an individual case, as they are found in two very different banks. 

% We also wonder if the engineer is given the first turn to come up with a form design, will the final product be different because of their habits in managing function reuses? We leave this question and the study as to how form designs could have affected the underlying implementations and occurrences of vulnerabilities to the future work.
\section{Related Work}
\label{sec:crs_related}

\subsection{Parameter Tampering Scanners} 
\textbf{Blackbox}. Blackbox scanners are not tied to any specific frameworks and languages used on the server-side. This makes them often more generic than whitebox approaches. In general, they all work by first crawling an application to capture any requests encountered, then fuzzing it by replaying the requests with some mutated parameters. 
Proxy-based parameter tampering scanners include the HTTP fuzzer in Acunetix~\cite{acunetix}, WebScarab~\cite{webscarab}, and TamperData~\cite{tamperdata}. They accept per-request fuzzing, and thus cannot automatically observe the intended workflow. They also require considerable amount of manual configurations such as specifying (a) a self-signed SSL certificate for interceptions of encrypted traffic~\cite{sslock}, and; (b) even explicit value range of each parameter.
 
NoTamper~\cite{notamper} outperforms proxy-based parameter tampering by reducing the manual efforts needed. Similar to CRS, it automates parameter generation based on the client-side validations. With static symbolic execution and constraint solving (that currently support the event-based form submissions only), it is able to generate an exhaustive set of parameters that are rejected by the client-side. This approach can complement the Parameter Mutator of CRS. It however requires an analyst to provide two (i.e., valid and invalid) initial sets of requests though, so as to discern whether the server accepts the mutated requests.

To preserve the intended workflow during fuzzing, research work is found only for scanning other web vulnerabilities such as XSS. A state-aware fuzzer~\cite{state-aware} is able to detect whether a fuzzing attempt has triggered a state change. If changed, it backtracks to the previous state by page traversal. It thus allows every state to be completely fuzzed by all candidate parameters before moving on to the next state. However, this stateful scanner can handle static web pages only, i.e., AJAX applications are not supported.

Nonetheless, all these scanners are limited owing to their fundamental scanner design. It is very hard to enhance them in handling the one-time use tokens, cross-request parameter dependencies, and client-side pre-processing. Their fuzzing requests are thus likely barred from reaching vulnerable code.

On the other hand, there are studies that evaluate the effectiveness of existing web vulnerability scanners. According to~\cite{blackboxeval}, the commercial scanners being evaluated are found not to be as comprehensive as they are claimed to be. This evaluation however covers only vulnerabilities like XSS, and it does not cover parameter tampering. According to~\cite{whyjohnny}, the study found that none of its evaluated scanners could find a parameter tampering vulnerability. It also suggests that high-level domain knowledge on the application is needed, but can hardly be known without a human.

PAPAS~\cite{parampollution} focuses only on a specific variant of parameter tampering, namely the parameter pollution vulnerabilities. Its success relies on the use of a server-side parameter precedence algorithm that allows malicious parameter to take over the precedence of an existing parameter. Its scanning approach is thus very similar to finding reflected vectors such as XSS, and that is different from finding more general parameter tampering vulnerabilities.

\textbf{Whitebox}. When server-side source code is available, whitebox scanners can be used to compare the inconsistencies between client-side and server-side validations.

WAPTEC~\cite{waptec} extends the NoTamper~\cite{notamper} approach by applying similar coding analysis to the server-side validations. The additional knowledge eliminates the manual effort required, and improves the selection of candidate parameter. But it still suffers from the limitations of NoTamper such as the negligence of the cross-request parameter dependency.

ViewPoints~\cite{viewpoints} discovers validation inconsistencies between the client-side and server-side by differential string analysis. It supports only a specific framework of the Java web applications. Besides reporting those parameters that are client-rejected yet server-accepted, its goal is to also report parameters that are client-accepted yet server-rejected. Eliminating this latter class of inconsistencies from the application can improve in usability but not its security.

In general, what can be validated from the server and client sides is intrinsically inconsistent since the server is always more knowledgeable (e.g., access to DB) than the client, and thus these approaches must report many false positives. It is also difficult to first annotate all the expected inconsistencies particularly in real-world complex applications before running these tools.

\subsection{Parameter Tampering Mitigations}

InteGuard~\cite{integuard} is a whitebox protection approach that operates a server-side proxy to verify the parameters exchanged between the server and a client as well as those further exchanged with a third-party server. It relies on a human to provide some valid transactions for analyzing the parameter dependencies in its learning phase. Then its online defense is to drop subsequent requests that violate the expected dependencies. Its mitigation is limited only to those parameters triggered by human. CRS can automatically explore the client-side events to uncover new parameters, which can be hardly exhausted by a human in the training phase.

Some development frameworks such as .NET~\cite{dotnet} and Java~\cite{googlewebtoolkit} can replicate the server-side validations to become client-side code. However, they are often limited to simple validations that are purposely programmed for replications. Swift~\cite{swift} proposes that developers should annotate those parameters and codes that are sensitive to run only on server-side or safe for both sides to facilitate replication. But again, it is demanding to annotate even a simple program.

Server-side validations and sanitizations are still fundamental defenses against parameter tampering. For client-side validations aiming at delivering better user experience, HTML 5 Validations~\cite{html5} and JavaScript libraries such as jQuery~\cite{jquery} can be deployed.

\subsection{Others}
Compared to~\cite{sso,shop-for-free} which also explored parameter relationships, CRS systematically analyzes all the parameter dependencies between different tuples of requests and responses, as shown in Figure~\ref{fig:arch}. As a result, CRS can automatically detect server acceptances by analyzing \texttt{Req vs Resp} and \texttt{Resp vs Resp'}, which makes a significant improvement over the manual and protocol-specific approaches as with~\cite{sso,shop-for-free}. The proposed model checking also requires human effort in accurately transforming a subset of logic replica from the original application~\cite{shop-for-free}.

Some existing work also involve in replaying user actions, but none of which have identified the advantage of fuzzing inside the application context, that is, by intercepting the submission requests for fuzzing. Mugshot~\cite{mugshot} deterministically captures and replays user actions and other JavaScript functions for failure and usability analysis. Kudzu~\cite{kudzu} enhances its code coverage by exploring client-side events for the discovery of more client-side injection vulnerabilities. Ripley~\cite{ripley} maintains a server-side replica of the client-side logic in a shadow browser hosted in a trusted proxy, and attempts to automatically replicate and replay the client-side events in the replica for integrity check. Although it ensures the computation integrity of the client-side code, it does not eliminate the need for server-side validations. Acunetix Web Vulnerability Scanner~\cite{acunetixwvs} is able to generate some user action events while scanning for XSS vulnerabilities.

\section{Conclusion}

We studied a number of real-world web applications to understand their implementations, and how they can be intercepted for vulnerability scanning. Existing parameter tampering scanners do not consider the enforcements of, among others, the intended workflow, one-time use tokens, and parameter dependency across requests, which are all common in multi-request applications. Their effectiveness of vulnerability scanning is thus severely limited. We proposed the novel CRS approach to respect all these enforcements. We realized the approach by building a generic blackbox parameter tampering scanner. In our evaluation, it is practical to drive real applications to run for fuzzing. The importance of this work is demonstrated by the vulnerabilities uncovered in real-world applications including banks, which existing scanners miss. The scanning approach can also be extended to the discovery of other web application vulnerabilities including XSS and CSRF.

\section*{Acknowledgements}
This work is partially supported by the Sir Edward Youde
Memorial Fellowship. We appreciate and thank HSBC and Hong Kong Monetary Authority
for their prompt follow-up. Special thanks are due to Chris Smith,
Clara Ho, Thomas Yung, Esmond Lee, Shu-pui Li, and James Tam
for their support. We thank for the valuable comments
from Shuo Chen, Kehuan Zhang and anonymous reviewers.

%
% The following two commands are all you need in the
% initial runs of your .tex file to
% produce the bibliography for the citations in your paper.
\bibliographystyle{abbrv}
\bibliography{references}  % sigproc.bib is the name of the Bibliography in this case

\begin{thebibliography}{10}

\bibitem{acunetix}
Acunetix.
\newblock Http fuzzer.
\newblock {\em \url{http://www.acunetix.com/blog/docs/http-fuzzer-tool/}}.

\bibitem{acunetixwvs}
Acunetix.
\newblock Web vulnerability scanner.
\newblock {\em \url{http://www.acunetix.com/vulnerability-scanner/}}.

\bibitem{viewpoints}
M.~Alkhalaf, S.~R. Choudhary, M.~Fazzini, T.~Bultan, A.~Orso, and C.~Kruegel.
\newblock Viewpoints: differential string analysis for discovering client- and
  server-side input validation inconsistencies.
\newblock In {\em Proceedings of the 2012 International Symposium on Software
  Testing and Analysis}, ISSTA 2012, pages 56--66, New York, NY, USA, 2012.
  ACM.

\bibitem{dotnet}
ASP.
\newblock Net validation controls.
\newblock {\em \url{http://msdn.microsoft.com/en-us/library/debza5t0.aspx}}.

\bibitem{parampollution}
M.~Balduzzi, C.~T. Gimenez, D.~Balzarotti, and E.~Kirda.
\newblock Automated discovery of parameter pollution vulnerabilities in web
  applications.
\newblock In {\em Proceedings of the 18th Annual Network and Distributed System
  Security Symposium}, NDSS'11, San Diego, CA, USA, 2011. The Internet Society.

\bibitem{blackboxeval}
J.~Bau, E.~Bursztein, D.~Gupta, and J.~Mitchell.
\newblock State of the art: Automated black-box web application vulnerability
  testing.
\newblock In {\em Proceedings of the 2010 IEEE Symposium on Security and
  Privacy}, SP '10, pages 332--345, Washington, DC, USA, 2010. IEEE Computer
  Society.

\bibitem{notamper}
P.~Bisht, T.~Hinrichs, N.~Skrupsky, R.~Bobrowicz, and V.~N. Venkatakrishnan.
\newblock Notamper: automatic blackbox detection of parameter tampering
  opportunities in web applications.
\newblock In {\em Proceedings of the 17th ACM conference on Computer and
  communications security}, CCS '10, pages 607--618, New York, NY, USA, 2010.
  ACM.

\bibitem{waptec}
P.~Bisht, T.~Hinrichs, N.~Skrupsky, and V.~N. Venkatakrishnan.
\newblock Waptec: whitebox analysis of web applications for parameter tampering
  exploit construction.
\newblock In {\em Proceedings of the 18th ACM conference on Computer and
  communications security}, CCS '11, pages 575--586, New York, NY, USA, 2011.
  ACM.

\bibitem{swift}
S.~Chong, J.~Liu, A.~C. Myers, X.~Qi, K.~Vikram, L.~Zheng, and X.~Zheng.
\newblock Secure web applications via automatic partitioning.
\newblock In {\em Proceedings of 21st ACM SIGOPS symposium on Operating systems
  principles}, SOSP '07, pages 31--44, New York, NY, USA, 2007. ACM.

\bibitem{immutable}
CWE.
\newblock Cwe-472: External control of assumed-immutable web parameter.
\newblock {\em \url{http://cwe.mitre.org/data/definitions/472.html}}.

\bibitem{state-aware}
A.~Doup{\'e}, L.~Cavedon, C.~Kruegel, and G.~Vigna.
\newblock Enemy of the state: a state-aware black-box web vulnerability
  scanner.
\newblock In {\em Proceedings of the 21st USENIX conference on Security
  symposium}, Security'12, pages 26--26, Berkeley, CA, USA, 2012. USENIX
  Association.

\bibitem{whyjohnny}
A.~Doup{\'e}, M.~Cova, and G.~Vigna.
\newblock Why johnny can't pentest: an analysis of black-box web vulnerability
  scanners.
\newblock In {\em Proceedings of the 7th international conference on Detection
  of intrusions and malware, and vulnerability assessment}, DIMVA'10, pages
  111--131, Berlin, Heidelberg, 2010. Springer-Verlag.

\bibitem{logicvul}
V.~Felmetsger, L.~Cavedon, C.~Kruegel, and G.~Vigna.
\newblock Toward automated detection of logic vulnerabilities in web
  applications.
\newblock In {\em Proceedings of the 19th USENIX conference on Security},
  USENIX Security'10, pages 10--10, Berkeley, CA, USA, 2010. USENIX
  Association.

\bibitem{sslock}
A.~P.~H. Fung and K.~W. Cheung.
\newblock Sslock: Sustaining the trust on entities brought by ssl.
\newblock In {\em Proceedings of the 5th ACM Symposium on Information, Computer
  and Communications Security}, ASIACCS '10, pages 204--213, New York, NY, USA,
  2010. ACM.

\bibitem{hsbcaward}
{Global Finance}.
\newblock World's best internet banks 2011.
\newblock {\em
  \url{http://www.gfmag.com/tools/best-banks/11485-worlds-best-internet-banks-2011.html}}.

\bibitem{googlewebtoolkit}
{Google Web Toolkit}.
\newblock Validation framework. concept of operations.
\newblock {\em
  \url{http://code.google.com/p/gwt-validation/wiki/ConceptOfOperations}}.

\bibitem{hsbcnewdevice}
HSBC.
\newblock Faq - security device.
\newblock {\em \url{https://www.hsbc.com.hk/1/2/hk/misc/otphelp}}.

\bibitem{html5}
{HTML 5}.
\newblock Association of controls and forms.
\newblock {\em
  \url{http://dev.w3.org/html5/spec/association-of-controls-and-forms.html}}.

\bibitem{json}
JSON.
\newblock Javascript objection notation.
\newblock {\em \url{http://json.org/}}.

\bibitem{tamperdata}
A.~Judson.
\newblock Tamper data :: Add-ons for firefox.
\newblock {\em
  \url{https://addons.mozilla.org/en-US/firefox/addon/tamper-data/}}.

\bibitem{mugshot}
J.~Mickens, J.~Elson, and J.~Howell.
\newblock Mugshot: deterministic capture and replay for javascript
  applications.
\newblock In {\em Proceedings of the 7th USENIX conference on Networked systems
  design and implementation}, NSDI'10, pages 11--11, Berkeley, CA, USA, 2010.
  USENIX Association.

\bibitem{xpath}
Mozilla.
\newblock Introduction to using xpath in javascript.
\newblock {\em
  \url{https://developer.mozilla.org/en/Introduction\_to\_using\_XPath\_in\_Javascript}}.

\bibitem{mutationobserver}
Mozilla.
\newblock Mutationobserver.
\newblock {\em
  \url{https://developer.mozilla.org/en-US/docs/DOM/MutationObserver}}.

\bibitem{approx-string}
G.~Navarro.
\newblock A guided tour to approximate string matching.
\newblock {\em ACM Computing Surveys}, 33(1):31--88, Mar. 2001.

\bibitem{webscarab}
OWASP.
\newblock Fuzzing with webscarab.
\newblock {\em \url{https://www.owasp.org/index.php/Fuzzing\_with\_WebScarab}}.

\bibitem{owasp-insecure-direct-object}
OWASP.
\newblock {Top 10 2013 A4-Insecure Direct Object References}.
\newblock {\em
  \url{https://www.owasp.org/index.php/Top\_10\_2013-A4-Insecure\_Direct\_Object\_References}}.

\bibitem{phantomjs}
PhantomJS.
\newblock Headless webkit with javascript api.
\newblock {\em \url{http://phantomjs.org/}}.

\bibitem{jquery}
J.~Resig.
\newblock jquery.
\newblock {\em \url{http://jquery.com/}}.

\bibitem{kudzu}
P.~Saxena, D.~Akhawe, S.~Hanna, F.~Mao, S.~McCamant, and D.~Song.
\newblock A symbolic execution framework for javascript.
\newblock In {\em Proceedings of the 2010 IEEE Symposium on Security and
  Privacy}, SP '10, pages 513--528, Washington, DC, USA, 2010. IEEE Computer
  Society.

\bibitem{selenium}
Selenium.
\newblock Web browser automation.
\newblock {\em \url{http://seleniumhq.org/}}.

\bibitem{skipfish}
Skipfish.
\newblock Web application security scanner.
\newblock {\em \url{https://code.google.com/p/skipfish/}}.

\bibitem{ripley}
K.~Vikram, A.~Prateek, and B.~Livshits.
\newblock Ripley: automatically securing web 2.0 applications through
  replicated execution.
\newblock In {\em Proceedings of the 16th ACM conference on Computer and
  communications security}, CCS '09, pages 173--186, New York, NY, USA, 2009.
  ACM.

\bibitem{sso}
R.~Wang, S.~Chen, and X.~Wang.
\newblock Signing me onto your accounts through facebook and google: A
  traffic-guided security study of commercially deployed single-sign-on web
  services.
\newblock In {\em Proceedings of the 2012 IEEE Symposium on Security and
  Privacy}, SP '12, pages 365--379, Washington, DC, USA, 2012. IEEE Computer
  Society.

\bibitem{shop-for-free}
R.~Wang, S.~Chen, X.~Wang, and S.~Qadeer.
\newblock How to shop for free online -- security analysis of
  cashier-as-a-service based web stores.
\newblock In {\em Proceedings of the 2011 IEEE Symposium on Security and
  Privacy}, SP '11, pages 465--480, Washington, DC, USA, 2011. IEEE Computer
  Society.

\bibitem{integuard}
L.~Xing, Y.~Chen, X.~Wang, and S.~Chen.
\newblock Integuard: Toward automatic protection of third-party web service
  integrations.
\newblock In {\em Proceedings of the 20th Annual Network and Distributed System
  Security Symposium}, NDSS'13, San Diego, CA, USA, 2013. The Internet Society.

\bibitem{keyword-highlight}
S.~Yang.
\newblock Search engine keyword highlighting with javascript.
\newblock {\em \url{http://scott.yang.id.au/code/se-hilite/}}.

\end{thebibliography}
\end{document}